\documentclass[pre,twocolumn,aps,floats,superscriptaddress,longbibliography,
floatfix,showpacs,showkeys,amsmath,amssymb,nofootinbib]{revtex4-1}

\usepackage{dcolumn}
\usepackage{bm}
\usepackage{txfonts}
\usepackage{pxfonts}
\usepackage{graphicx}
\usepackage{psfrag}
\usepackage{times}

\begin{document}
\title{Surface tension and instability in the hydrodynamic white hole of 
a circular hydraulic jump}

\author{Jayanta K. Bhattacharjee}\email{jayanta.bhattacharjee@gmail.com}
\affiliation{School of Physical Sciences, Indian Association
for the Cultivation of Science, Jadavpur, Kolkata 700032, India} 

\author{Arnab K. Ray}\email{arnab_kumar@daiict.ac.in}
\affiliation{Dhirubhai Ambani Institute of Information and
Communication Technology, Gandhinagar 382007, Gujarat, India}

\date{\today}

\begin{abstract}
We impose a linearized Eulerian perturbation on a steady, shallow, 
radial outflow of a liquid (water), whose local pressure function 
includes both the hydrostatic and the Laplace pressure terms.
The resulting wave equation bears the form of a
hydrodynamic metric. A dispersion relation, extracted from the wave
equation, gives an instability due to surface tension and the 
cylindrical flow symmetry. Using the dispersion relation, we also
derive three known relations that scale the radius of the circular 
hydraulic jump in the outflow. The first two relations
are scaled by viscosity and gravity, with a capillarity-dependent 
crossover to the third relation, which is scaled by viscosity and 
surface tension. The perturbation as a high-frequency travelling 
wave, 
propagating radially inward against the bulk outflow, is blocked 
just outside 
the circular hydraulic jump. The amplitude of the wave also diverges
here because of a singularity.
The blocking is associated with surface tension, which renders the 
circular hydraulic jump a hydrodynamic white hole. 
\end{abstract}

\pacs{47.20.Dr, 47.35.Bb, 47.35.Pq, 04.80.Cc} 
\keywords{Surface-tension-driven instability; Gravity waves; 
Capillary waves; Experimental tests of gravitational theories}

\maketitle

\section{Introduction} 
\label{sec1}
The hydraulic jump is 
an abrupt discontinuity in the free-surface height of a flowing 
liquid, with the post-jump height being greater than the 
pre-jump height~\citep{ll87}. The circular hydraulic jump is 
formed by the outward radial flow of a liquid on a horizontal 
plane, after the flow originates at the point of impingement of 
a vertically downward liquid jet~\citep{tan49}. The flow in this
case is commonly viewed as shallow and axially symmetric. The formation
of a hydraulic jump depends on two defining attributes of a 
liquid, namely, viscosity and surface tension. Various theories 
have addressed the question of how the circular hydraulic jump 
forms because of  
viscosity~\citep{tan49,watson,bdp93,bpw97,rb07,kas08} and surface 
tension~\citep{ba03,rat13,dab19}, with experimental evidence in
favour of both  
means~\citep{tan49,ot66,behh96,hansen97,ba03,bah06,rgp07,kdc07,
dll14,bjlw18,aaekpr19}.

Apart from conventional fluid dynamics, the hydraulic 
jump is viewed with keen interest from the perspective of the 
fluid analogue of gravity, specifically as a hydrodynamic white 
hole~\citep{su02,vol05,vol06,rb07,jpmmr,blv11,jkb17}. In this respect, 
the circular hydraulic jump is like a standing event horizon 
where the velocity of the radial flow equals the local speed 
of either surface gravity waves~\citep{rb07} or capillary-gravity 
waves~\citep{vol05}. The hydrodynamic event horizon 
segregates the supercritical and the subcritical regions
of the flow, where the critical condition refers to the matching 
of the speed of the bulk flow with the speed of the waves. The 
horizon (which spatially coincides with 
the circular jump) thus outlines a barrier against the upstream 
transmission of information, i.e. becomes a hydrodynamic white hole. 
As the equatorial flow proceeds outwards from its point of origin,
its radial velocity is greater than the speed of gravity
waves, but viscosity and the radial geometry decelerate
the flow downstream. When the critical condition is met, 
the jump and the horizon occur simultaneously~\citep{rb07}.
The horizon of a white hole in the circular hydraulic 
jump has been analyzed in theoretical 
studies~\citep{vol05,rb07,jkb17}
and demonstrated experimentally~\citep{jpmmr}. However, 
the horizon of the white hole by itself is inadequate to explain 
why a jump should coincide with it~\citep{jkb17}. The jump is 
brought about by energy dissipation (at the discontinuity), 
which is a general
principle, established by~\citet{jws14} (also see~\citep{ll87}).
In shallow laboratory flows of normal liquids, mainly viscosity 
causes the jump at 
positions of the centimetre order~\citep{watson,bdp93,sbr05}. 
For jump positions of smaller length scales, surface tension 
is the main cause~\citep{ba03}, as is known for 
submicron molten metal droplets~\citep{rama07} and 
superfluids~\citep{rgp07}.
In the present work, we revisit the 
circular hydraulic jump and the associated hydrodynamic horizon
from the viewpoint of surface tension.

The outflow that we study here pertains to 
the standard Type-I hydraulic jump, which is 
physically characterized by a negligible flow height at the 
outer boundary (the perimeter of the horizontal flow base), 
where the liquid falls freely off the edge of the base
plane~\citep{bdp93,behh96}. In Sec.~\ref{sec2} we set down 
the height-averaged equations of such a flow, taken to be 
shallow. Its local 
pressure function accounts for the effect of both hydrostatics
and surface tension. 
An Eulerian perturbation of the steady flow establishes the 
wave equation of a fluid metric in Sec.~\ref{sec3}. 
We derive a dispersion relation from the wave equation in 
Sec.~\ref{sec4}. The dispersion relation brings forth two salient
results. The first is an instability in the flow because of 
surface tension and the cylindrical symmetry.
Secondly, the dispersion relation unifies three known 
scaling formulae of the radius of the circular hydraulic 
jump in a single theoretical framework. The first two formulae
depend on viscosity and 
gravity~\citep{bdp93,rat13} and the third depends on 
viscosity and surface tension~\citep{bjlw18}. Viscosity and
gravity determine the jump scaling when the wavelength of 
the perturbation far exceeds the capillary length, whereas 
surface tension becomes a dominant effect at the cost of 
gravity when the capillary length is much greater than the
wavelength of the perturbation. The crossover from
the one regime to the other occurs when the capillary length
and the wavelength are evenly matched.  

Lastly, in Sec.~\ref{sec5}, on applying the {\it WKB} approximation, 
we design the perturbation to be a travelling wave of high frequency.
By this we show that the hydrodynamic event horizon,
where the hydraulic jump is also located, blocks a wave 
that travels from the subcritical flow region towards the
circular jump, against the bulk outflow. The amplitude of 
this travelling wave diverges as the jump radius is approached,
thereby demonstrating that the jump is like a white-hole 
to any approaching wave from the subcritical region. 
To support our theory, we provide photographic evidence
from an experiment by~\citet{kdc07}. 

\section{The height-averaged flow equations} 
\label{sec2} 
Cylindrical coordinates, 
$(r,\phi,z)$, are suitable for a shallow radial flow 
of liquid on a plane~\citep{ll87}. The flow, being axially symmetric, 
is independent of the azimuthal coordinate, $\phi$.  
Being also shallow, the flow allows a vertical averaging of the flow 
variables~\citep{bdp93,bpw97} through the height of the flow, 
under the boundary conditions that
velocities vanish at $z=0$ (the no-slip condition), and vertical
gradients of velocities vanish at the free surface of the flow
(the no-stress condition)~\citep{bdp93,bpw97,sbr05,kas08}. 
The boundary conditions are applied under the operative assumption 
that the vertical component of the velocity is small 
compared to its radial component, and the vertical variation of 
the radial velocity (through the shallow layer of fluid) is much 
greater than its radial variation~\citep{bdp93}.
Quantities with the $z$-coordinate are averaged thus through 
the flow height, 
with the double $z$-derivative approximated as
$\partial^2/\partial z^2 \equiv -1/h^2$~\citep{bdp93}, 
in which $h$ is the free-surface height of the flow. 

In terms of $h$ and the vertically-averaged radial velocity, 
$v$, the time-dependent continuity equation of the shallow-water 
circular flow is~\citep{rb07,rsbb18} 
\begin{equation} 
\label{conteq} 
\frac{\partial h}{\partial t} + \frac{1}{r} 
\frac{\partial}{\partial r}
\left(rvh \right) =0.  
\end{equation} 
Likewise, the Navier-Stokes equation of the time-dependent 
radial component of the flow is~\citep{rb07,rsbb18}  
\begin{equation} 
\label{avgns} 
\frac{\partial v}{\partial t} + v\frac{\partial v}{\partial r} 
+\frac{1}{\rho}\frac{\partial P}{\partial r}= -\frac{\nu v}{h^2},
\end{equation} 
whose right hand side has been approximated by
$\nu \nabla^2 v\simeq -\nu v/h^2$, for a thin layer of the
flow~\citep{bdp93}. Here $\nu$ is the kinematic viscosity. 
The solutions of Eqs.~(\ref{conteq}) and~(\ref{avgns}), 
$h(r,t)$ and $v(r,t)$, can be known upon prescribing a function 
for the pressure, $P$, in Eq.~(\ref{avgns}). Since we are 
concerned with 
the effect of surface tension, in addition to the usual 
hydrostatic pressure, we account for the surface pressure,
as given by Laplace's formula~\citep{ll87,ba03,dab19}. 
Together, these two effects give   
\begin{equation} 
\label{press} 
P= h \rho g - \frac{\sigma}{r}\frac{\partial}{\partial r} 
\left[\frac{r}{\sqrt{1+(\partial h/\partial r)^2}}
\frac{\partial h}{\partial r}\right], 
\end{equation} 
with the first term on the right hand side being the hydrostatic
pressure, containing the liquid density, $\rho$, and 
the acceleration due to gravity, $g$. The second term on the 
right hand side of Eq.~(\ref{press}) is the contribution that 
surface tension, $\sigma$, makes to the pressure. With $P$ 
set down in terms of $h$ and $r$, the coupled system of 
Eqs.~(\ref{conteq}) and~(\ref{avgns}) is now closed. 

In the steady state of the shallow radial flow, 
whereby $\partial/\partial t \equiv 0$, the solutions of 
Eqs.~(\ref{conteq}) and~(\ref{avgns}) are $h_0(r)$ and $v_0(r)$, 
with the subscript ``$0$" standing for steady values. The integral 
of Eq.~(\ref{conteq}) in the steady limit gives $rv_0h_0 = Q/2\pi$, 
in which $Q$ is the steady volumetric flow rate. In the absence 
of surface tension ($\sigma =0$), the behaviour of the steady 
solutions and the conditions to form the circular hydraulic jump
are known~\citep{bdp93,rb07}. The hydraulic jump is a discontinuity 
in the solutions of $h_0(r)$ and $v_0(r)$. From the perspective
of fluid analogues of gravity, an event horizon exists at the 
radius of the hydraulic jump, $r=r_{\mathrm J}$, where
$v_0^2(r_{\mathrm J})=gh_0(r_{\mathrm J})$~\citep{rb07}. 
Hereafter, we proceed to examine how the fluid event horizon and 
the hydraulic jump are affected by surface tension. 

\section{Perturbation and a hydrodynamic metric}
\label{sec3}
We define a variable, $f=rvh$, in which $v \equiv v(r,t)$
and $h \equiv h(r,t)$ for the vertically-averaged radial 
flow~\citep{rb07,rsbb18}. Under steady
conditions, $f=f_0=Q/2\pi$, as Eq.~(\ref{conteq}) shows. 
Next, we apply a time-dependent radial perturbation
on the flow that Eqs.~(\ref{conteq}) and~(\ref{avgns}) describe. 
The perturbations in $v$ and $h$ are set, respectively, as
$v(r,t)=v_0(r)+v^{\prime}(r,t)$ and $h(r,t)=h_0(r)+h^{\prime}(r,t)$.
With $f^\prime$ denoting a perturbation in $f$, we linearize 
according to $f=f_0+f^\prime$ and get  
\begin{equation}
\label{radf} 
f^\prime =r\left(v_0h^\prime + h_0 v^\prime \right). 
\end{equation} 
Under this Eulerian perturbation scheme, 
the fluctuation about the steady continuity condition is 
obtained from Eq.~(\ref{conteq}) as~\citep{rb07} 
\begin{equation}
\label{rconf} 
\frac{\partial h^\prime}{\partial t} = -\frac{1}{r} 
\frac{\partial f^\prime}{\partial r}, 
\end{equation} 
which relates $h^\prime$ to $f^\prime$. A similar relation, 
connecting $v^\prime$ and $f^\prime$, is then derived by 
using Eq.~(\ref{rconf}) in Eq.~(\ref{radf}). This is~\citep{rb07}
\begin{equation} 
\label{effvee} 
\frac{\partial v^\prime}{\partial t} = \frac{v_0}{f_0} 
\left(\frac{\partial f^\prime}{\partial t} +v_0 
\frac{\partial f^\prime}{\partial r}\right).
\end{equation} 

Now we perturb $v$ and $h$ in Eq.~(\ref{avgns}) to a linear order
about the steady state. Then taking the time derivative of the 
resulting linearized equation, and applying both 
Eqs.~(\ref{rconf}) and~(\ref{effvee}) in it, we finally get 
a wave equation
\begin{multline}
\label{rmetric} 
\frac{\partial}{\partial t}\left(v_0
\frac{\partial f^\prime}{\partial t}\right)
+\frac{\partial}{\partial t}\left(v_0^2
\frac{\partial f^\prime}{\partial r}\right)
+\frac{\partial}{\partial r}\left(v_0^2
\frac{\partial f^\prime}{\partial t}\right) \\
+\frac{\partial}{\partial r}\left[v_0
\left(v_0^2-gh_0+\frac{gh_0l^2\Gamma}{r^2}\right)
\frac{\partial f^\prime}{\partial r}\right] 
=-\frac{\nu v_0}{h_0^2}\left(\frac{\partial f^\prime}{\partial t} 
+3v_0\frac{\partial f^\prime}{\partial r}\right) \\
-v_0gh_0rl^2\frac{\partial}{\partial r}\left[
\frac{1}{r\{1+({\mathrm d}h_0/{\mathrm d}r)^2\}^{3/2}}
\frac{\partial^3 f^\prime}{\partial r^3}-\frac{\Gamma}{r^2}
\frac{\partial^2 f^\prime}{\partial r^2}\right], 
\end{multline}
in which $l$ is the capillary length, 
$l=\sqrt{\sigma/(\rho g)}$~\citep{ll87}, and
\begin{equation}
\label{defgam} 
\Gamma =\frac{1}{[1+({\mathrm d}h_0/{\mathrm d}r)^2]^{3/2}}
\left[1+\frac{3{\mathrm d}h_0/{\mathrm d}r}
{1+({\mathrm d}h_0/{\mathrm d}r)^2}
\frac{{\mathrm d}^2h_0}{{\mathrm d}r^2}\right]. 
\end{equation}
The capillary length, $l$, introduced in Eq.~(\ref{rmetric}), 
captures the effect of surface tension. The 
importance of surface tension can be gauged by comparing 
the capillary length with the wavelength of the perturbation. 
We show this in Sec.~\ref{sec4}. 

If the terms depending on viscosity and surface tension 
in Eq.~(\ref{rmetric}) are to vanish 
($\nu =0$ and $\sigma=0$), then the resulting wave equation 
in $f^\prime$ is rendered compactly as
$\partial_\alpha \left( {\mathsf{f}}^{\alpha \beta}
\partial_\beta f^{\prime}\right) =0$,
in which the Greek indices run from $0$ to $1$, with
$0$ standing for $t$ and $1$ standing for $r$. 
From the terms on the left hand side of Eq.~(\ref{rmetric}),
we can then read the symmetric matrix,
\begin{equation}
\label{symmat}
{\mathsf{f}}^{\alpha \beta } = v_0
\begin{bmatrix}
1 & v_0 \\
v_0 & v_0^2 - gh_0
\end{bmatrix} \\. 
\end{equation}
The basic principle of 
establishing a hydrodynamic metric and an analogue horizon
rests on an equivalence between Eq.~(\ref{symmat}) and the 
d'Alembertian for a scalar field in curved geometry 
(see~\citep{rb07} and relevant references therein for details).
The d'Alembertian has the form~\citep{blv11}
\begin{equation}
\label{alem}
\triangle \psi \equiv \frac{1}{\sqrt{-\mathsf{g}}}
\partial_\alpha \left({\sqrt{-\mathsf{g}}}\, {\mathsf{g}}^{\alpha \beta} 
\partial_\beta \psi \right).
\end{equation}
Under the identification that ${\mathsf{f}}^{\alpha \beta } =
\sqrt{-\mathsf{g}}\, {\mathsf{g}}^{\alpha \beta}$, and therefore,
$\mathsf{g} = \det \left({\mathsf{f}}^{\alpha \beta }\right)$, 
one can prove the existence of a white hole horizon for surface
waves when $v_0^2 = gh_0$~\citep{su02,rb07}. This produces a 
fluid analogue of a general relativistic problem~\citep{su02}. 

By disregarding both viscosity and surface tension in a normal 
liquid, 
the symmetry of the metric implied 
by Eq.~(\ref{symmat}) may be preserved, but the absence of  
viscosity (and surface tension) will also 
prevent the formation of a physical inner solution 
within the jump radius~\citep{bdp93}. This difficulty is overcome  
by accounting for viscosity in the radial outflow~\citep{bdp93}, 
even though it has to invalidate the hydrodynamic metric. The precise 
condition of the analogue horizon is thus lost. Nevertheless, the basic
properties of surface waves are not affected overmuch~\citep{su02}, 
and the most important feature to emerge from the analogy of a
white hole horizon remains qualitatively unchanged, namely, that
a disturbance propagating upstream from the subcritical flow region
(where $v_0^2 < gh_0$) cannot penetrate through the horizon into the
supercritical region of the flow (where $v_0^2 > gh_0$), both in 
the presence of viscosity~\citep{rb07} and surface 
tension~\citep{vol05}. 

\section{Dispersion, instability and jump scaling}
\label{sec4}
Looking closely at Eq.~(\ref{rmetric}), we discern in it the
form of the wave equation. Consequently, we can extract a 
dispersion relation from it. An approximation makes the 
dispersion relation stand out clearly. We know that the 
steady free-surface height of the flow varies slowly 
with $r$ for the greater part of the flow. Therefore, we approximate 
${\mathrm d}h_0/{\mathrm d}r \simeq 0$, which gives 
$\Gamma \simeq 1$ and simplifies Eq.~(\ref{rmetric}) to
\begin{multline} 
\label{difeqgam1}
\frac{\partial^2 f^\prime}{\partial t^2}+2\frac{\partial}{\partial r} 
\left(v_0 \frac{\partial f^\prime}{\partial t}\right) + \frac{1}{v_0} 
\frac{\partial}{\partial r}\left[v_0\left(v_0^2-gh_0\right)
\frac{\partial f^\prime}{\partial r}\right]= \\
-\frac{\nu}{h_0^2}
\left(\frac{\partial f^\prime}{\partial t} 
+3v_0\frac{\partial f^\prime}{\partial r}\right) 
-\frac{\sigma h_0r}{\rho}\frac{\partial}{\partial r}
\left(\frac{1}{r}\frac{\partial^3 f^\prime}{\partial r^3}
-\frac{1}{r^2}\frac{\partial^2 f^\prime}{\partial r^2} 
+\frac{1}{r^3}\frac{\partial f^\prime}{\partial r}\right). 
\end{multline} 
With respect to the steady background flow, Eq.~(\ref{difeqgam1}) 
becomes
\begin{multline} 
\label{waveq} 
\frac{\partial^2 f^\prime}{\partial t^2}=gh_0 
\frac{\partial^2 f^\prime}{\partial r^2}
-\frac{\nu}{h_0^2}\frac{\partial f^\prime}{\partial t} \\
-\frac{\sigma h_0}{\rho}\left(\frac{\partial^4 f^\prime}{\partial r^4}
-\frac{2}{r}\frac{\partial^3 f^\prime}{\partial r^3}
+\frac{3}{r^2}\frac{\partial^2 f^\prime}{\partial r^2}
-\frac{3}{r^3}\frac{\partial f^\prime}{\partial r}\right),
\end{multline} 
which is the wave equation (for gravity waves)
when $\nu =0$ and $\sigma =0$. The solution,
$f^\prime (r,t) \sim \exp [i(kr-\omega t)]$,
applied to Eq.~(\ref{waveq}), gives a quadratic equation,
\begin{multline} 
\label{disquad} 
\left(\omega -kv_{\mathrm B}\right)^2=\left(gh_0-
\frac{3\sigma h_0}{\rho r^2}\right)k^2 + \frac{\sigma h_0 k^4}{\rho} \\
-\frac{i\nu}{h_0^2}\left(\omega -kv_{\mathrm B}\right)
+\frac{i\sigma h_0}{\rho}\left(\frac{2k^3}{r}-\frac{3k}{r^3}\right),
\end{multline} 
in which $v_{\mathrm B}$ stands for the bulk motion of the liquid.
The outcome due to viscosity is well known and has been reported 
in detail previously~\citep{rb07}. Hence, we set $\nu =0$ in our 
present study and devote our attention fully to the effect of 
$\sigma$ in the wave equation. This gives us  
\begin{equation} 
\label{disper0} 
\omega =kv_{\mathrm B} +\sqrt{gh_0}\left[1-\frac{3l^2}{r^2}+l^2k^2
+\frac{il^2}{r^2}\left(2kr-\frac{3}{kr}\right)\right]^{1/2} k.
\end{equation}
In the reference frame of $v_{\mathrm B}=0$, we can 
view Eq.~(\ref{disper0}) in the form,  
$\omega = \sqrt{gh_0}\left(A + iB\right)^{1/2}k 
= \sqrt{gh_0}\left(X+iY\right)k$, in which $A$, $B$, $X$ and $Y$ 
are all real quantities. These are related among themselves by 
$X=\pm \left[\left(\sqrt{A^2+B^2}+A\right)/2\right]^{1/2}$ 
and $Y=\pm \left[\left(\sqrt{A^2+B^2}-A\right)/2\right]^{1/2}$. 

In Eq.~(\ref{disper0}), the terms with $l^2/r^2$ have arisen 
because of the cylindrical symmetry of the shallow flow. For 
water, $l=0.27\,\mathrm{cm}$ and $r \sim 10\,\mathrm{cm}$, which 
makes $l^2/r^2 \ll 1$. Since $B$ contains only a single term 
with $l^2/r^2$ and $A$ is at least $\mathcal{O}(1)$, as we see 
in Eq.~(\ref{disper0}), it is clear that $B \ll A$. Hence, by
a binomial expansion, we can approximate 
$Y \simeq \pm B/\left(2\sqrt{A}\right)$. Since both signs are 
admissible, with $A \sim 1$ and $B \sim l^2/r^2$, we realize 
that the amplitude of $f^\prime$ can grow as 
\begin{equation} 
\label{ampligrow} 
{\big{\vert}}f^\prime (r,t) {\big{\vert}} \sim \exp 
\left(\frac{l^2 k\sqrt{gh_0}t}{r^2}\right),
\end{equation} 
on a time scale of $r^2/\left(l^2 k \sqrt{gh_0}\right)$. 
Since $r^2 \gg l^2$, this is a long time scale for the growth
of an instability in the flow. Apropos of this, surface 
ripples~\citep{ot66}, capillary-gravity waves~\citep{hansen97} 
and instability due to surface tension~\citep{hansen97,kas08} 
are known for normal liquids. Similar features have also been
observed in superfluids~\citep{rgp07}, which we stress here 
because we have set $\nu =0$.

Going back to Eq.~(\ref{disper0}) in the reference frame of 
$v_{\mathrm B}=0$, and neglecting $l^2/r^2$ in it, we get a 
comoving dispersion relation, 
\begin{equation} 
\label{disper} 
\omega \simeq \sqrt{gh_0} \left(1+l^2k^2\right)^{1/2} k. 
\end{equation} 
The foregoing equation is actually the long-wavelength limiting
case of the dispersion relation for capillary-gravity waves,   
$\omega^2 = \left[gk+\left(\sigma/\rho\right)k^3\right]
\tanh\left(kh_0\right)$, for $kh_0 \ll 1$ (which approximates 
to $\tanh \left(kh_0\right) \simeq kh_0$)~\citep{ll87}. 
In the limit of $k \ll h_0^{-1}$, 
the wavelength, $\lambda \gg h_0$, which is 
the case of long wavelengths in shallow-water flows. 
Thus, this condition is implicit in Eq.~(\ref{disper}) and all 
the equations that lead to it, starting with 
Eqs.~(\ref{conteq}) and~(\ref{avgns}). 

From Eq.~(\ref{disper}) we derive some familiar scaling 
relations for the radius of the circular hydraulic jump.
First, for $kl \ll 1$, i.e.
$\lambda \gg l$, Eq.~(\ref{disper}) gives the phase velocity 
of gravity waves, 
$v_{\mathrm p} =\omega/k \simeq \sqrt{gh_0}$. 
Now, viscosity affects the bulk motion, which is seen 
by comparing the first term on the left hand side of 
Eq.~(\ref{avgns}) with the viscosity term on the right hand side. 
The time scale on which viscosity decelerates 
the flow is $t_{\mathrm{visc}}\sim h_0^2/\nu$~\citep{rb07}.  
The deceleration of an advanced layer of the flow by viscosity 
can be known upstream
if a travelling wave carries the information against the
flow. But this information travels at the speed of surface gravity 
waves, $\sqrt{gh_0}$, something that is known from Eq.~(\ref{disper})
when $kl \ll 1$. Therefore, the supercritical region, where 
$v_0 > \sqrt{gh_0}$, remains uninformed about the viscous 
deceleration downstream~\citep{rb07}. The flow thus proceeds 
radially outwards without hindrance in the supercritical region 
till $v_0$ becomes comparable with $\sqrt{gh_0}$, and 
only then does the information about an obstacle ahead 
catch up with the fluid. Defining a dynamic time scale,
$t_{\mathrm{dyn}}\sim r/v_0$ (the time scale of  
the bulk motion) and setting 
$t_\mathrm{visc} \simeq t_\mathrm{dyn}$, with the additional
requirements, $v_0 \simeq v_{\mathrm p} \simeq \sqrt{gh_0}$ and 
$rv_0h_0=Q/2\pi$, 
deliver a scale of the jump radius as
\begin{equation}
\label{bdpscal}
r_\mathrm{J} \sim Q^{5/8} \nu^{-3/8} g^{-1/8},
\end{equation} 
due originally to~\citet{bdp93}, with some 
later refinements~\citep{dll14}. To derive the scaling relation in 
Eq.~(\ref{bdpscal}),~\citet{bdp93} matched steady inner and outer 
solutions through a standing shock at the hydraulic jump. On 
the other hand,  
the crux of our chain of reasoning to arrive at Eq.~(\ref{bdpscal}) 
is that the hydraulic jump forms when the two time scales, 
$t_{\mathrm{visc}}$ and $t_{\mathrm{dyn}}$, 
match each other closely, and when the Froude number, 
$\mathcal{F}=v_0/\sqrt{gh_0} \simeq 1$. In matching the 
time scales, $t_{\mathrm{visc}}$ brings in viscosity as 
a physical means for the jump to form, for which the 
unity of $\mathcal{F}$ (implying only 
the horizon) is not enough~\citep{jkb17}. 
Under the combined effect of 
all these  conditions, a layer of fluid arriving late is
halted by an obstacle formed by a layer of fluid ahead, 
slowed abruptly by viscosity. But the
fluid cannot accumulate indefinitely, and also the continuity 
of the fluid flow is to be maintained. The fluid layer 
arriving late, therefore, slides over the slowly flowing
layer ahead, causing a sudden increase in the flow height 
--- a hydraulic jump~\citep{rb07}.

Once a scaling relation is known for $r_\mathrm{J}$, as in 
Eq.~(\ref{bdpscal}), the height of the jump is then scaled as
$h_\mathrm{J} \sim Q^{1/4} \nu^{1/4} g^{-1/4}$. In the post-jump
region of the flow, this height remains nearly unchanged, which 
lets the post-jump flow height, $H$, to be set as 
$H \sim h_\mathrm{J}$. 
Now, Eq.~(\ref{bdpscal}) 
scales $r_\mathrm{J}$ in terms of the free parameters of the 
flow, $Q$, $\nu$ and $g$, but in terms of $H$, 
Eq.~(\ref{bdpscal}) can be recast into a scaling relation, 
\begin{equation}
\label{rojscal}
r_\mathrm{J} \sim Q^{3/4} \nu^{-1/4} g^{-1/4} H^{-1/2}, 
\end{equation}
due to~\citet{rat13}. As opposed to Eq.~(\ref{bdpscal}), the
jump radius in Eq.~(\ref{rojscal}) is not scaled by the
free parameters of the flow, but by the depth of the liquid
downstream of the jump. 

The basic premise of Eqs.~(\ref{bdpscal}) and~(\ref{rojscal}),
both free of the surface tension, $\sigma$,  
is that $kl \ll 1$ in Eq.~(\ref{disper}). In the opposite limit 
of $kl \gg 1$, Eq.~(\ref{disper}) approximates to 
$\omega \simeq \sqrt{(\sigma h_0)/\rho}k^2$. Forcing the 
condition, $k \sim h_0^{-1}$, one gets the phase velocity, 
$v_\mathrm{p} = \omega/k \simeq \sqrt{\sigma/(\rho h_0)}$. 
Thereafter, using the condition 
$v_0 \simeq v_\mathrm{p} \simeq \sqrt{\sigma/(\rho h_0)}$
and following the same line of reasoning that led to 
Eq.~(\ref{bdpscal}), one arrives at a scaling relation, 
\begin{equation}
\label{bhagscal}
r_\mathrm{J} \sim Q^{3/4} \rho^{1/4} \nu^{-1/4} \sigma^{-1/4},
\end{equation}
due to~\citet{bjlw18}. The noteworthy aspect of Eq.~(\ref{bhagscal}) 
is that it is free of gravity, $g$, but depends on the surface 
tension, $\sigma$. The scaling formula of Eq.~(\ref{bhagscal}) 
can also be derived by forcing the condition, 
$H \simeq l$, 
in Eq.~(\ref{rojscal})~\citep{bjlw18}.   

Looking at Eq.~(\ref{disper}), we realize that 
Eqs.~(\ref{bdpscal}) and~(\ref{rojscal}), both dependent on gravity
but free of surface tension, are valid in the limit
of $kl \ll 1$ (or $\lambda \gg l$). In contrast, Eq.~(\ref{bhagscal}),
free of gravity but dependent on surface tension, is 
valid in the opposite limit of $kl \gg 1$ (or $\lambda \ll l$). 
The crossover from the former regime to the latter happens when 
$\lambda \sim l$. The different scaling relations here show that 
a shallow flow can generally accommodate
the effects of both gravity and capillarity~\citep{aaekpr19}. 

Our derivation of the scaling formulae 
in Eqs.~(\ref{bdpscal}),~(\ref{rojscal})
and~(\ref{bhagscal}) is based on the matching of viscous and 
dynamical time scales when the flow becomes critical. An alternative 
approach to the same end is through a first-order dynamical system in
the steady flow. A relevant critical point of this dynamical system 
is a spiral~\citep{tan49,bdp93,kas08}. However, a physical
flow must be single-valued and cannot have a spiral profile. 
Therefore, close to 
the spiral critical point, an inner solution is matched to an outer 
solution through a jump discontinuity~\citep{bdp93,kas08}. 
The inner and outer solutions 
are uniquely characterized by the inner and outer boundary conditions, 
respectively~\citep{kas08}. Moreover, the matching of 
these solutions at the jump implies that its location is also   
determined by the boundary conditions. This is seen in both 
radial flows~\citep{kas08} and channel flows~\citep{sbr05}. 

\section{Wave blocking at the analogue horizon}
\label{sec5}
We subject the liquid outflow to a high-frequency travelling wave
under the {\it WKB} approximation~\citep{rb07}. When both $\nu =0$ 
and $\sigma =0$, the travelling wave does not destabilize 
the flow~\citep{rb07}. However, just outside the event horizon, 
viscosity causes a large divergence in the amplitude of the  
wave that propagates against the outward bulk flow~\citep{rb07}. 
Since the destabilizing effect of viscosity is known 
already~\citep{rb07}, we ignore the viscosity-dependent terms
(i.e. set $\nu =0$) in Eq.~(\ref{difeqgam1}). Thereafter, what 
remains of Eq.~(\ref{difeqgam1}) is subjected to a solution of  
the form, $f^\prime (r,t)= p(r)\exp(-i\omega t)$, which leads to
\begin{multline}
\label{peewkb}
\left(v_0^2-gh_0\right)\frac{{\mathrm d}^2 p}{{\mathrm d}r^2} 
+\left[\frac{1}{v_0}\frac{\mathrm d}{{\mathrm d}r}\left(v_0^3
-v_0gh_0\right)-2i\omega v_0\right]\frac{{\mathrm d}p}{{\mathrm d}r}\\
-\left(\omega^2+2i\omega \frac{{\mathrm d}v_0}{{\mathrm d}r}\right)p
= -\frac{\sigma h_0}{\rho}\left(
\frac{{\mathrm d}^4 p}{{\mathrm d}r^4} 
-\frac{2}{r}\frac{{\mathrm d}^3 p}{{\mathrm d}r^3} 
+\frac{3}{r^2}\frac{{\mathrm d}^2 p}{{\mathrm d}r^2} 
-\frac{3}{r^3}\frac{{\mathrm d}p}{{\mathrm d}r}\right). 
\end{multline}
For the spatial part of the solution, we prescribe $p(r)=e^s$, 
with $s\equiv s(r)$ given by a converging power series~\citep{rb07}, 
\begin{equation} 
\label{powser} 
s(r)=\sum_{n=-1}^\infty \frac{\tilde{k}_n(r)}{\omega^n}. 
\end{equation}
The convergence is ensured if the frequency, $\omega$, is high, 
so that any term in the power series of Eq.~(\ref{powser}) 
becomes much smaller than its preceding term, i.e. 
$\omega^{-(n+1)}\tilde{k}_{n+1} \ll \omega^{-n}\tilde{k}_n$. This 
condition is physically satisfied if the wavelength is smaller
than a characteristic length scale of the flow, which, in this 
instance, is the jump radius itself. Hence, 
under the {\it WKB} approximation, only the first two terms are 
significant, with the former contributing to the phase of the 
travelling perturbation and the latter to its amplitude~\citep{rb07}. 

With surface tension included, the highest derivative in 
Eq.~(\ref{peewkb}) is of the quartic order. In applying the 
{\it WKB} approximation, we, therefore, adopt an iterative approach. 
First, we set $\sigma =0$, and write all $\tilde{k}_n$ in $s(r)$ 
as $k_n$, with the latter implying the solution series of $s(r)$ 
without surface tension. Then, accounting for the first two terms
in the series of $s(r)$, along with the approximation that 
$\omega k_{-1} \gg k_0$, we gather all the coefficients of 
$\omega^2$ (the highest order of $\omega$) to arrive at~\citep{rb07}
\begin{equation} 
\label{kayminus} 
k_{-1}= i \int \frac{1}{v_0 \mp \sqrt{gh_0}}\,{\mathrm d}r.
\end{equation} 
Solving likewise for the coefficients of $\omega$ gives us
\begin{equation} 
\label{kaynot} 
k_0 =-\frac{1}{2} \ln \left(v_0\sqrt{gh_0}\right) +C,
\end{equation} 
in which $C$ is an integration constant. 
The convergence of $s(r) \simeq \omega k_{-1}+ k_0$ can be 
verified self-consistently from Eqs.~(\ref{kayminus}) 
and~(\ref{kaynot}) by showing that $\omega k_{-1} \gg k_0$~\citep{rb07}. 

Now we take up Eq.~(\ref{peewkb}) with $\sigma \neq 0$, and 
in it we apply $s(r)$ as given in Eq.~(\ref{powser}). The highest
order of $\omega$ in terms that are explicitly without $\sigma$ 
is $\omega^2$, and the highest order of $\omega$ in terms that 
explicitly have $\sigma$ is $\omega^4$. Gathering the former from
the left hand side and the latter from the right hand side gives
\begin{equation} 
\label{quadkayminus}
\left(v_0^2-gh_0\right)
\left(\frac{{\mathrm d}\tilde{k}_{-1}}{{\mathrm d}r}\right)^2 
-2iv_0 \frac{{\mathrm d}\tilde{k}_{-1}}{{\mathrm d}r}-1 \simeq 
-\frac{\sigma h_0}{\rho}\left(
\frac{{\mathrm d}k_{-1}}{{\mathrm d}r}\right)^4 \omega^2.
\end{equation} 
In our iterative approach we have approximated 
$\tilde{k}_{-1} \simeq k_{-1}$ on the right hand side of 
Eq.~(\ref{quadkayminus}), where $\sigma$ is explicitly present.  
This approximation is valid for small values of $\sigma$, whereby 
the capillary length, $l$, will be much smaller than the wavelength
of the travelling perturbation in the shallow-water flow. 
Solving the quadratic form of 
${\mathrm d}\tilde{k}_{-1}/{\mathrm d}r$ in Eq.~(\ref{quadkayminus}), 
we get 
\begin{equation} 
\label{kayminuscor} 
\tilde{k}_{-1} \simeq k_{-1}\pm i \int
\frac{\omega^2 l^2 \sqrt{gh_0}}{2(v_0 \mp \sqrt{gh_0})^4}\,{\mathrm d}r.
\end{equation} 
The second term on the right hand side of Eq.~(\ref{kayminuscor}) 
adds a surface-tension-dependent correction to what we already know 
from Eq.~(\ref{kayminus}). This correction is of the order of $\omega^2$,
and appears to be dominant over $k_{-1}$. 
This, however, is not really the case. Noting that the wavelength, 
$\lambda (r)=2\pi(v_0 \mp \sqrt{gh_0})/\omega$, we immediately see 
that the correction term in Eq.~(\ref{kayminuscor}) is subdominant 
to $k_{-1}$, when $l \ll \lambda$. This validates our iterative 
method self-consistently.  

After $\omega^2$, the next order is of $\omega$ in all the terms 
that are explicitly free of $\sigma$, while terms that explicitly 
bear $\sigma$ come with $\omega^3$ as the next higher order, 
following $\omega^4$. Terms with $\omega$ on the left hand side 
and $\omega^3$ on the right hand side lead to 
\begin{multline} 
\label{quadkaynot} 
2\left[\left(v_0^2-gh_0\right)
\frac{{\mathrm d}\tilde{k}_{-1}}{{\mathrm d}r}-iv_0\right]
\frac{{\mathrm d}\tilde{k}_0}{{\mathrm d}r}+\frac{1}{v_0} 
\frac{\mathrm d}{{\mathrm d}r}\left[v_0\left(v_0^2-gh_0\right)
\frac{{\mathrm d}\tilde{k}_{-1}}{{\mathrm d}r}\right] \\
-2i\frac{{\mathrm d}v_0}{{\mathrm d}r} \simeq 
-\frac{\sigma \omega^2 h_0}{\rho}\Bigg[
4\left(\frac{{\mathrm d}k_{-1}}{{\mathrm d}r}\right)^3
\frac{{\mathrm d}k_0}{{\mathrm d}r}+
6\left(\frac{{\mathrm d}k_{-1}}{{\mathrm d}r}\right)^2
\frac{{\mathrm d}^2k_{-1}}{{\mathrm d}r^2} \\
-\frac{2}{r}
\left(\frac{{\mathrm d}k_{-1}}{{\mathrm d}r}\right)^3
\Bigg] \simeq \frac{2\sigma \omega^2 h_0}{\rho r}
\left(\frac{{\mathrm d}k_{-1}}{{\mathrm d}r}\right)^3,  
\end{multline} 
in which, on the right hand side, we ultimately retain only the 
term that is most significant. Adopting the same line of 
reasoning, as has been done following Eq.~(\ref{quadkayminus}), 
gives
\begin{equation}
\label{kaynotcor} 
\tilde{k}_0 \simeq k_0 \mp \int
\frac{\omega^2 l^2}{gh_0r({\mathcal F} \mp 1)^3}\,{\mathrm d}r.
\end{equation} 
The second term on the right hand side of Eq.~(\ref{kaynotcor})
adds a correction to $k_0$, as given in Eq.~(\ref{kaynot}). We 
stress once again that $\omega^2 l^2$ renders the correction term 
subdominant to $k_0$. 

In the travelling perturbation, which can now be written as 
$f^\prime (r,t) \simeq \exp(\omega \tilde{k}_{-1}+\tilde{k_0} 
-i\omega t)$, we see that $\tilde{k}_{-1}$ contributes to the 
phase and $\tilde{k}_0$ contributes to the amplitude. Since we
are concerned with the stability of the travelling wave, we extract 
its amplitude, which, expressed in full, is 
\begin{equation} 
\label{fampli} 
\big{\vert} f^\prime (r,t) \big{\vert} \sim 
\left(v_0 \sqrt{gh_0}\right)^{-1/2} \exp \left[\mp \int
\frac{\omega^2 l^2}{gh_0r({\mathcal F} \mp 1)^3}\,{\mathrm d}r\right].
\end{equation} 
The upper sign in Eq.~(\ref{fampli}) pertains to a wave that 
propagates upstream against the outward radial bulk flow of liquid. 
We look at this case closely. 
In the subcritical region of the flow, where ${\mathcal F} <1$, the
integrand in Eq.~(\ref{fampli}) is negative. As the wave approaches
the singularity, which is owed entirely to gravity and where 
${\mathcal F}=1$, the integrand diverges. With the 
negative sign outside the integral, the overall outcome is 
$\vert f^\prime (r,t) \vert \longrightarrow \infty$, 
i.e. the wave suffers an instability. The very opposite of all
this occurs just inside the singularity. Here, with ${\mathcal F} >1$,
the integral acquires a negative sign overall, which results in 
$\vert f^\prime (r,t) \vert \longrightarrow 0$. 
This discontinuity in the inward propagation of the wave is 
forced by the term with surface tension in Eq.~(\ref{fampli}).
Since this happens at the analogue event horizon, we can say that 
the horizon acts like an impenetrable barrier (a fluid analogue
of a white hole) 
against incoming waves from the subcritical region. 

\begin{figure}
\begin{center}
\includegraphics[scale=1.3, angle=0]{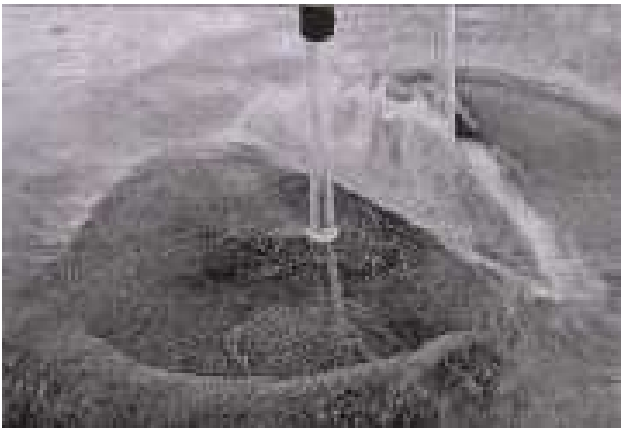}
\caption{\label{f1}\small{An oblique view of a steady arch-shaped 
upwash fountain 
between two 
circular hydraulic jumps. The close approach of one jump 
to the other is like a collision of two white holes. 
The photograph is by the courtesy of R. P. Kate, P. K. Das and 
S. Chakraborty~\citep{kdc07}.}}
\end{center}
\end{figure}

The foregoing theoretical claim receives support from an experiment 
carried out by~\citet{kdc07}. The experiment was 
on the interaction of two adjacent hydraulic jumps formed by normally 
impinging water jets, of which one was static and the other was 
mobile~\citep{kdc07}.\footnote{
This experiment~\citep{kdc07} supported a theory of the formation 
of circular hydraulic jumps due to viscosity~\citep{rb07}. 
Since surface tension has as much of a role to play as viscosity
to form circular hydraulic jumps, we refer to 
the same experiment in support of our present study.
In a qualitative sense, both viscosity and surface tension 
are responsible for blocking waves against 
the bulk flow at the event horizon.}
The photograph in Fig.\ref{f1}, taken by~\citet{kdc07}, shows 
clearly that when one water jet is moved 
close to the other one, the water trapped along the 
stagnation line between the 
circular hydraulic jumps created by the two jets is raised to a 
greater height than the rest of the flow. A steady arch-like upwash 
fountain 
thus comes to stand by itself (like a standing wall of water). The jump 
formed by the moving water jet is like a subcritical disturbance
propagating upstream towards the jump formed due to the static jet.
This disturbance is blocked by the static jump, which is 
an unyielding fluid white hole. Consequently, as the disturbance
approaches the static circular jump, the free-surface height of 
the water increases dramatically due to the accumulation of water. 
This agrees with what we have concluded from Eq.~(\ref{fampli}), 
namely, the blocking of a wave approaching the singularity from the 
subcritical region. Furthermore, our conjecture is that the steady 
upwash fountain between the two contiguous hydraulic jumps could 
be the fluid analogue of the compression and bulging of the spacetime 
geometry between two colliding general relativistic white holes.  
As a caveat we point out that in
the experiment of~\citet{kdc07}, the disturbance propagating 
upstream is not axisymmetric about the static jump, and so the 
standing wall of water between the two jumps is not axisymmetric 
either.

The derivation of the time-averaged 
energy flux of the perturbation in a two-dimensional radial flow 
has been established in a previous study~\citep{rb07}. By   
the same method, energy fluctuations of the 
first-order disappear upon time averaging, but second-order 
terms survive to contribute to the time-averaged energy flux, $F$.
With this contribution, we can show that 
$F \sim \langle{\vert f^\prime (r,t)\vert}^2\rangle$. Since 
$\vert f^\prime (r,t)\vert$ diverges just outside the analogue 
event horizon 
for a wave propagating against the radial outflow, $F$ will also 
exhibit a similar divergence about the same spatial location~\citep{rb07}. 

\section{Concluding remarks}
\label{sec6}
This theoretical study on the effect of surface tension in 
Type-I hydraulic jumps
has revealed two types of instabilities. One, as in 
Eq.~(\ref{ampligrow}), results from surface tension and the cylindrical
geometry of the shallow flow. The other, as in Eq.~(\ref{fampli}), 
is the combined outcome of gravity and surface tension. Gravity waves
define the location of the singularity in 
Eq.~(\ref{fampli}), but the divergence just outside the horizon 
singularity occurs because of surface tension. 
Surface tension is known to cause other instabilities as well.
For instance, the breaking of the axial symmetry of the steady circular 
hydraulic jump is an instability for which surface tension is 
responsible~\citep{bah06,kas08}.  

The scaling formula proposed by~\citet{bjlw18}, as in 
Eq.~(\ref{bhagscal}), has been the subject of close scrutiny because
of its exclusion of gravity~\citep{dab19,aaekpr19}. However, such
scaling has been argued to be valid for developing hydraulic jumps
in the capillary regime~\citep{aaekpr19}. In any case, it is known
that surface tension is much more significant than gravity
for circular jumps of small radii~\citep{ba03}. Femtocups created 
through gravity-free hydraulic jumps of molten metals are a case 
in point~\citep{rama07}. 

\begin{acknowledgments}
AKR gratefully acknowledges the hospitality of the Indian 
Association for the Cultivation of Science, Kolkata, India.  
\end{acknowledgments}

\bibliography{arXbr092021} 

\begin{thebibliography}{29}%
\makeatletter
\providecommand \@ifxundefined [1]{%
 \@ifx{#1\undefined}
}%
\providecommand \@ifnum [1]{%
 \ifnum #1\expandafter \@firstoftwo
 \else \expandafter \@secondoftwo
 \fi
}%
\providecommand \@ifx [1]{%
 \ifx #1\expandafter \@firstoftwo
 \else \expandafter \@secondoftwo
 \fi
}%
\providecommand \natexlab [1]{#1}%
\providecommand \enquote  [1]{``#1''}%
\providecommand \bibnamefont  [1]{#1}%
\providecommand \bibfnamefont [1]{#1}%
\providecommand \citenamefont [1]{#1}%
\providecommand \href@noop [0]{\@secondoftwo}%
\providecommand \href [0]{\begingroup \@sanitize@url \@href}%
\providecommand \@href[1]{\@@startlink{#1}\@@href}%
\providecommand \@@href[1]{\endgroup#1\@@endlink}%
\providecommand \@sanitize@url [0]{\catcode `\\12\catcode `\$12\catcode
  `\&12\catcode `\#12\catcode `\^12\catcode `\_12\catcode `\%12\relax}%
\providecommand \@@startlink[1]{}%
\providecommand \@@endlink[0]{}%
\providecommand \url  [0]{\begingroup\@sanitize@url \@url }%
\providecommand \@url [1]{\endgroup\@href {#1}{\urlprefix }}%
\providecommand \urlprefix  [0]{URL }%
\providecommand \Eprint [0]{\href }%
\providecommand \doibase [0]{http://dx.doi.org/}%
\providecommand \selectlanguage [0]{\@gobble}%
\providecommand \bibinfo  [0]{\@secondoftwo}%
\providecommand \bibfield  [0]{\@secondoftwo}%
\providecommand \translation [1]{[#1]}%
\providecommand \BibitemOpen [0]{}%
\providecommand \bibitemStop [0]{}%
\providecommand \bibitemNoStop [0]{.\EOS\space}%
\providecommand \EOS [0]{\spacefactor3000\relax}%
\providecommand \BibitemShut  [1]{\csname bibitem#1\endcsname}%
\let\auto@bib@innerbib\@empty
\bibitem [{\citenamefont {Landau}\ and\ \citenamefont {Lifshitz}(1987)}]{ll87}%
  \BibitemOpen
  \bibfield  {author} {\bibinfo {author} {\bibfnamefont {L.~D.}\ \bibnamefont
  {Landau}}\ and\ \bibinfo {author} {\bibfnamefont {E.~M.}\ \bibnamefont
  {Lifshitz}},\ }\href@noop {} {\emph {\bibinfo {title} {Course of Theoretical
  Physics: Fluid Mechanics}}}\ (\bibinfo  {publisher} {Butterworth-Heinemann},\
  \bibinfo {address} {Oxford},\ \bibinfo {year} {1987})\BibitemShut {NoStop}%
\bibitem [{\citenamefont {Tani}(1949)}]{tan49}%
  \BibitemOpen
  \bibfield  {author} {\bibinfo {author} {\bibfnamefont {I.}~\bibnamefont
  {Tani}},\ }\bibfield  {title} {\enquote {\bibinfo {title} {Water jump in the
  boundary layer},}\ }\href@noop {} {\bibfield  {journal} {\bibinfo  {journal}
  {J. Phys. Soc. Japan}\ }\textbf {\bibinfo {volume} {4}},\ \bibinfo {pages}
  {212} (\bibinfo {year} {1949})}\BibitemShut {NoStop}%
\bibitem [{\citenamefont {Watson}(1964)}]{watson}%
  \BibitemOpen
  \bibfield  {author} {\bibinfo {author} {\bibfnamefont {E.~J.}\ \bibnamefont
  {Watson}},\ }\bibfield  {title} {\enquote {\bibinfo {title} {The radial
  spread of a liquid jet over a horizontal plane},}\ }\href@noop {} {\bibfield
  {journal} {\bibinfo  {journal} {J. Fluid Mech.}\ }\textbf {\bibinfo {volume}
  {20}},\ \bibinfo {pages} {481} (\bibinfo {year} {1964})}\BibitemShut
  {NoStop}%
\bibitem [{\citenamefont {Bohr}\ \emph {et~al.}(1993)\citenamefont {Bohr},
  \citenamefont {Dimon},\ and\ \citenamefont {Putkaradze}}]{bdp93}%
  \BibitemOpen
  \bibfield  {author} {\bibinfo {author} {\bibfnamefont {T.}~\bibnamefont
  {Bohr}}, \bibinfo {author} {\bibfnamefont {P.}~\bibnamefont {Dimon}}, \ and\
  \bibinfo {author} {\bibfnamefont {V.}~\bibnamefont {Putkaradze}},\ }\bibfield
   {title} {\enquote {\bibinfo {title} {Shallow-water approach to the circular
  hydraulic jump},}\ }\href@noop {} {\bibfield  {journal} {\bibinfo  {journal}
  {J. Fluid Mech.}\ }\textbf {\bibinfo {volume} {254}},\ \bibinfo {pages} {635}
  (\bibinfo {year} {1993})}\BibitemShut {NoStop}%
\bibitem [{\citenamefont {Bohr}\ \emph {et~al.}(1997)\citenamefont {Bohr},
  \citenamefont {Putkaradze},\ and\ \citenamefont {Watanabe}}]{bpw97}%
  \BibitemOpen
  \bibfield  {author} {\bibinfo {author} {\bibfnamefont {T.}~\bibnamefont
  {Bohr}}, \bibinfo {author} {\bibfnamefont {V.}~\bibnamefont {Putkaradze}}, \
  and\ \bibinfo {author} {\bibfnamefont {S.}~\bibnamefont {Watanabe}},\
  }\bibfield  {title} {\enquote {\bibinfo {title} {Averaging theory for the
  structure of hydraulic jumps and separation in laminar free-surface flows},}\
  }\href@noop {} {\bibfield  {journal} {\bibinfo  {journal} {Phys. Rev. Lett.}\
  }\textbf {\bibinfo {volume} {79}},\ \bibinfo {pages} {1038} (\bibinfo {year}
  {1997})}\BibitemShut {NoStop}%
\bibitem [{\citenamefont {Ray}\ and\ \citenamefont
  {Bhattacharjee}(2007)}]{rb07}%
  \BibitemOpen
  \bibfield  {author} {\bibinfo {author} {\bibfnamefont {A.~K.}\ \bibnamefont
  {Ray}}\ and\ \bibinfo {author} {\bibfnamefont {J.~K.}\ \bibnamefont
  {Bhattacharjee}},\ }\bibfield  {title} {\enquote {\bibinfo {title} {Standing
  and travelling waves in the shallow-water circular hydraulic jump},}\
  }\href@noop {} {\bibfield  {journal} {\bibinfo  {journal} {Phys. Lett. A}\
  }\textbf {\bibinfo {volume} {371}},\ \bibinfo {pages} {241} (\bibinfo {year}
  {2007})}\BibitemShut {NoStop}%
\bibitem [{\citenamefont {Kasimov}(2008)}]{kas08}%
  \BibitemOpen
  \bibfield  {author} {\bibinfo {author} {\bibfnamefont {A.~R.}\ \bibnamefont
  {Kasimov}},\ }\bibfield  {title} {\enquote {\bibinfo {title} {A stationary
  circular hydraulic jump, the limits of its existence and its gasdynamic
  analogue},}\ }\href@noop {} {\bibfield  {journal} {\bibinfo  {journal} {J.
  Fluid Mech.}\ }\textbf {\bibinfo {volume} {601}},\ \bibinfo {pages} {189}
  (\bibinfo {year} {2008})}\BibitemShut {NoStop}%
\bibitem [{\citenamefont {Bush}\ and\ \citenamefont {Aristoff}(2003)}]{ba03}%
  \BibitemOpen
  \bibfield  {author} {\bibinfo {author} {\bibfnamefont {J.~W.~M.}\
  \bibnamefont {Bush}}\ and\ \bibinfo {author} {\bibfnamefont {J.~M.}\
  \bibnamefont {Aristoff}},\ }\bibfield  {title} {\enquote {\bibinfo {title}
  {The influence of surface tension on the circular hydraulic jump},}\
  }\href@noop {} {\bibfield  {journal} {\bibinfo  {journal} {J. Fluid Mech.}\
  }\textbf {\bibinfo {volume} {489}},\ \bibinfo {pages} {229} (\bibinfo {year}
  {2003})}\BibitemShut {NoStop}%
\bibitem [{\citenamefont {Rojas}\ \emph {et~al.}(2013)\citenamefont {Rojas},
  \citenamefont {Argentina},\ and\ \citenamefont {Tirapegui}}]{rat13}%
  \BibitemOpen
  \bibfield  {author} {\bibinfo {author} {\bibfnamefont {N.}~\bibnamefont
  {Rojas}}, \bibinfo {author} {\bibfnamefont {M.}~\bibnamefont {Argentina}}, \
  and\ \bibinfo {author} {\bibfnamefont {E.}~\bibnamefont {Tirapegui}},\
  }\bibfield  {title} {\enquote {\bibinfo {title} {A progressive correction to
  the circular hydraulic jump scaling},}\ }\href@noop {} {\bibfield  {journal}
  {\bibinfo  {journal} {Physics of Fluids}\ }\textbf {\bibinfo {volume} {25}},\
  \bibinfo {pages} {042105} (\bibinfo {year} {2013})}\BibitemShut {NoStop}%
\bibitem [{\citenamefont {Duchesne}\ \emph {et~al.}(2019)\citenamefont
  {Duchesne}, \citenamefont {Andersen},\ and\ \citenamefont {Bohr}}]{dab19}%
  \BibitemOpen
  \bibfield  {author} {\bibinfo {author} {\bibfnamefont {A.}~\bibnamefont
  {Duchesne}}, \bibinfo {author} {\bibfnamefont {A.}~\bibnamefont {Andersen}},
  \ and\ \bibinfo {author} {\bibfnamefont {T.}~\bibnamefont {Bohr}},\
  }\bibfield  {title} {\enquote {\bibinfo {title} {Surface tension and the
  origin of the circular hydraulic jump in a thin liquid film},}\ }\href@noop
  {} {\bibfield  {journal} {\bibinfo  {journal} {Phys. Rev. Fluids}\ }\textbf
  {\bibinfo {volume} {4}},\ \bibinfo {pages} {084001} (\bibinfo {year}
  {2019})}\BibitemShut {NoStop}%
\bibitem [{\citenamefont {Olsson}\ and\ \citenamefont
  {Turkdogan}(1966)}]{ot66}%
  \BibitemOpen
  \bibfield  {author} {\bibinfo {author} {\bibfnamefont {R.~G.}\ \bibnamefont
  {Olsson}}\ and\ \bibinfo {author} {\bibfnamefont {E.~T.}\ \bibnamefont
  {Turkdogan}},\ }\bibfield  {title} {\enquote {\bibinfo {title} {Radial spread
  of a liquid stream on a horizontal plate},}\ }\href@noop {} {\bibfield
  {journal} {\bibinfo  {journal} {Nature}\ }\textbf {\bibinfo {volume} {211}},\
  \bibinfo {pages} {813} (\bibinfo {year} {1966})}\BibitemShut {NoStop}%
\bibitem [{\citenamefont {Bohr}\ \emph {et~al.}(1996)\citenamefont {Bohr},
  \citenamefont {Ellegaard}, \citenamefont {Hansen},\ and\ \citenamefont
  {Hanning}}]{behh96}%
  \BibitemOpen
  \bibfield  {author} {\bibinfo {author} {\bibfnamefont {T.}~\bibnamefont
  {Bohr}}, \bibinfo {author} {\bibfnamefont {C.}~\bibnamefont {Ellegaard}},
  \bibinfo {author} {\bibfnamefont {A.~E.}\ \bibnamefont {Hansen}}, \ and\
  \bibinfo {author} {\bibfnamefont {A.}~\bibnamefont {Hanning}},\ }\bibfield
  {title} {\enquote {\bibinfo {title} {Hydraulic jumps, flow separation and
  wave breaking: An experimental study},}\ }\href@noop {} {\bibfield  {journal}
  {\bibinfo  {journal} {Physica B}\ }\textbf {\bibinfo {volume} {228}},\
  \bibinfo {pages} {1} (\bibinfo {year} {1996})}\BibitemShut {NoStop}%
\bibitem [{\citenamefont {Hansen}\ \emph {et~al.}(1997)\citenamefont {Hansen},
  \citenamefont {H{\o}rl{\"u}ck}, \citenamefont {Zauner}, \citenamefont
  {Dimon}, \citenamefont {Ellegaard},\ and\ \citenamefont {Creagh}}]{hansen97}%
  \BibitemOpen
  \bibfield  {author} {\bibinfo {author} {\bibfnamefont {S.~H.}\ \bibnamefont
  {Hansen}}, \bibinfo {author} {\bibfnamefont {S.}~\bibnamefont
  {H{\o}rl{\"u}ck}}, \bibinfo {author} {\bibfnamefont {D.}~\bibnamefont
  {Zauner}}, \bibinfo {author} {\bibfnamefont {P.}~\bibnamefont {Dimon}},
  \bibinfo {author} {\bibfnamefont {C.}~\bibnamefont {Ellegaard}}, \ and\
  \bibinfo {author} {\bibfnamefont {S.~C.}\ \bibnamefont {Creagh}},\ }\bibfield
   {title} {\enquote {\bibinfo {title} {Geometric orbits of surface waves from
  a circular hydraulic jump},}\ }\href@noop {} {\bibfield  {journal} {\bibinfo
  {journal} {Phys. Rev. E}\ }\textbf {\bibinfo {volume} {55}},\ \bibinfo
  {pages} {7048} (\bibinfo {year} {1997})}\BibitemShut {NoStop}%
\bibitem [{\citenamefont {Bush}\ \emph {et~al.}(2006)\citenamefont {Bush},
  \citenamefont {Aristoff},\ and\ \citenamefont {Hosoi}}]{bah06}%
  \BibitemOpen
  \bibfield  {author} {\bibinfo {author} {\bibfnamefont {J.~W.~M.}\
  \bibnamefont {Bush}}, \bibinfo {author} {\bibfnamefont {J.~M.}\ \bibnamefont
  {Aristoff}}, \ and\ \bibinfo {author} {\bibfnamefont {A.~E.}\ \bibnamefont
  {Hosoi}},\ }\bibfield  {title} {\enquote {\bibinfo {title} {An experimental
  investigation of the stability of the circular hydraulic jump},}\ }\href@noop
  {} {\bibfield  {journal} {\bibinfo  {journal} {J. Fluid Mech.}\ }\textbf
  {\bibinfo {volume} {558}},\ \bibinfo {pages} {33} (\bibinfo {year}
  {2006})}\BibitemShut {NoStop}%
\bibitem [{\citenamefont {Rolley}\ \emph {et~al.}(2007)\citenamefont {Rolley},
  \citenamefont {Guthmann},\ and\ \citenamefont {Pettersen}}]{rgp07}%
  \BibitemOpen
  \bibfield  {author} {\bibinfo {author} {\bibfnamefont {E.}~\bibnamefont
  {Rolley}}, \bibinfo {author} {\bibfnamefont {C.}~\bibnamefont {Guthmann}}, \
  and\ \bibinfo {author} {\bibfnamefont {M.~S.}\ \bibnamefont {Pettersen}},\
  }\bibfield  {title} {\enquote {\bibinfo {title} {The hydraulic jump and
  ripples in liquid helium},}\ }\href@noop {} {\bibfield  {journal} {\bibinfo
  {journal} {Physica B}\ }\textbf {\bibinfo {volume} {394}},\ \bibinfo {pages}
  {46} (\bibinfo {year} {2007})}\BibitemShut {NoStop}%
\bibitem [{\citenamefont {Kate}\ \emph {et~al.}(2007)\citenamefont {Kate},
  \citenamefont {Das},\ and\ \citenamefont {Chakraborty}}]{kdc07}%
  \BibitemOpen
  \bibfield  {author} {\bibinfo {author} {\bibfnamefont {R.~P.}\ \bibnamefont
  {Kate}}, \bibinfo {author} {\bibfnamefont {P.~K.}\ \bibnamefont {Das}}, \
  and\ \bibinfo {author} {\bibfnamefont {S.}~\bibnamefont {Chakraborty}},\
  }\bibfield  {title} {\enquote {\bibinfo {title} {An experimental
  investigation on the interaction of hydraulic jumps formed by two normal
  impinging circular liquid jets},}\ }\href@noop {} {\bibfield  {journal}
  {\bibinfo  {journal} {J. Fluid Mech.}\ }\textbf {\bibinfo {volume} {590}},\
  \bibinfo {pages} {355} (\bibinfo {year} {2007})}\BibitemShut {NoStop}%
\bibitem [{\citenamefont {Duchesne}\ \emph {et~al.}(2014)\citenamefont
  {Duchesne}, \citenamefont {Lebon},\ and\ \citenamefont {Limat}}]{dll14}%
  \BibitemOpen
  \bibfield  {author} {\bibinfo {author} {\bibfnamefont {A.}~\bibnamefont
  {Duchesne}}, \bibinfo {author} {\bibfnamefont {L.}~\bibnamefont {Lebon}}, \
  and\ \bibinfo {author} {\bibfnamefont {L.}~\bibnamefont {Limat}},\ }\bibfield
   {title} {\enquote {\bibinfo {title} {Constant {F}roude number in a circular
  hydraulic jump and its implication on the jump radius selection},}\
  }\href@noop {} {\bibfield  {journal} {\bibinfo  {journal} {Europhys. Lett.}\
  }\textbf {\bibinfo {volume} {107}},\ \bibinfo {pages} {54002} (\bibinfo
  {year} {2014})}\BibitemShut {NoStop}%
\bibitem [{\citenamefont {Bhagat}\ \emph {et~al.}(2018)\citenamefont {Bhagat},
  \citenamefont {Jha}, \citenamefont {Linden},\ and\ \citenamefont
  {Wilson}}]{bjlw18}%
  \BibitemOpen
  \bibfield  {author} {\bibinfo {author} {\bibfnamefont {R.~K.}\ \bibnamefont
  {Bhagat}}, \bibinfo {author} {\bibfnamefont {N.~K.}\ \bibnamefont {Jha}},
  \bibinfo {author} {\bibfnamefont {P.~F.}\ \bibnamefont {Linden}}, \ and\
  \bibinfo {author} {\bibfnamefont {D.~I.}\ \bibnamefont {Wilson}},\ }\bibfield
   {title} {\enquote {\bibinfo {title} {On the origin of the circular hydraulic
  jump in a thin liquid film},}\ }\href@noop {} {\bibfield  {journal} {\bibinfo
   {journal} {J. Fluid Mech.}\ }\textbf {\bibinfo {volume} {851}},\ \bibinfo
  {pages} {R5} (\bibinfo {year} {2018})}\BibitemShut {NoStop}%
\bibitem [{\citenamefont {Askarizadeh}\ \emph {et~al.}(2019)\citenamefont
  {Askarizadeh}, \citenamefont {Ahmadikia}, \citenamefont {Ehrenpreis},
  \citenamefont {Kneer}, \citenamefont {Pishevar},\ and\ \citenamefont
  {Rohlfs}}]{aaekpr19}%
  \BibitemOpen
  \bibfield  {author} {\bibinfo {author} {\bibfnamefont {H.}~\bibnamefont
  {Askarizadeh}}, \bibinfo {author} {\bibfnamefont {H.}~\bibnamefont
  {Ahmadikia}}, \bibinfo {author} {\bibfnamefont {C.}~\bibnamefont
  {Ehrenpreis}}, \bibinfo {author} {\bibfnamefont {R.}~\bibnamefont {Kneer}},
  \bibinfo {author} {\bibfnamefont {A.}~\bibnamefont {Pishevar}}, \ and\
  \bibinfo {author} {\bibfnamefont {W.}~\bibnamefont {Rohlfs}},\ }\bibfield
  {title} {\enquote {\bibinfo {title} {Role of gravity and capillary waves in
  the origin of circular hydraulic jumps},}\ }\href@noop {} {\bibfield
  {journal} {\bibinfo  {journal} {Phys. Rev. Fluids}\ }\textbf {\bibinfo
  {volume} {4}},\ \bibinfo {pages} {114002} (\bibinfo {year}
  {2019})}\BibitemShut {NoStop}%
\bibitem [{\citenamefont {Sch{\"u}tzhold}\ and\ \citenamefont
  {Unruh}(2002)}]{su02}%
  \BibitemOpen
  \bibfield  {author} {\bibinfo {author} {\bibfnamefont {R.}~\bibnamefont
  {Sch{\"u}tzhold}}\ and\ \bibinfo {author} {\bibfnamefont {W.~G.}\
  \bibnamefont {Unruh}},\ }\bibfield  {title} {\enquote {\bibinfo {title}
  {Gravity wave analogues of black holes},}\ }\href@noop {} {\bibfield
  {journal} {\bibinfo  {journal} {Phys. Rev. D}\ }\textbf {\bibinfo {volume}
  {66}},\ \bibinfo {pages} {044019} (\bibinfo {year} {2002})}\BibitemShut
  {NoStop}%
\bibitem [{\citenamefont {Volovik}(2005)}]{vol05}%
  \BibitemOpen
  \bibfield  {author} {\bibinfo {author} {\bibfnamefont {G.~E.}\ \bibnamefont
  {Volovik}},\ }\bibfield  {title} {\enquote {\bibinfo {title} {Hydraulic jump
  as a white hole},}\ }\href@noop {} {\bibfield  {journal} {\bibinfo  {journal}
  {JETP Letters}\ }\textbf {\bibinfo {volume} {82}},\ \bibinfo {pages} {624}
  (\bibinfo {year} {2005})}\BibitemShut {NoStop}%
\bibitem [{\citenamefont {Volovik}(2006)}]{vol06}%
  \BibitemOpen
  \bibfield  {author} {\bibinfo {author} {\bibfnamefont {G.~E.}\ \bibnamefont
  {Volovik}},\ }\bibfield  {title} {\enquote {\bibinfo {title} {Horizons and
  ergoregions in superfluids},}\ }\href@noop {} {\bibfield  {journal} {\bibinfo
   {journal} {J. Low Temp. Phys.}\ }\textbf {\bibinfo {volume} {145}},\
  \bibinfo {pages} {337} (\bibinfo {year} {2006})}\BibitemShut {NoStop}%
\bibitem [{\citenamefont {Jannes}\ \emph {et~al.}(2011)\citenamefont {Jannes},
  \citenamefont {Piquet}, \citenamefont {Ma{\"i}ssa}, \citenamefont {Mathis},\
  and\ \citenamefont {Rousseaux}}]{jpmmr}%
  \BibitemOpen
  \bibfield  {author} {\bibinfo {author} {\bibfnamefont {G.}~\bibnamefont
  {Jannes}}, \bibinfo {author} {\bibfnamefont {R.}~\bibnamefont {Piquet}},
  \bibinfo {author} {\bibfnamefont {P.}~\bibnamefont {Ma{\"i}ssa}}, \bibinfo
  {author} {\bibfnamefont {C.}~\bibnamefont {Mathis}}, \ and\ \bibinfo {author}
  {\bibfnamefont {G.}~\bibnamefont {Rousseaux}},\ }\bibfield  {title} {\enquote
  {\bibinfo {title} {Experimental demonstration of the supersonic-subsonic
  bifurcation in the circular jump: A hydrodynamic white hole},}\ }\href@noop
  {} {\bibfield  {journal} {\bibinfo  {journal} {Phys. Rev. E}\ }\textbf
  {\bibinfo {volume} {83}},\ \bibinfo {pages} {056312} (\bibinfo {year}
  {2011})}\BibitemShut {NoStop}%
\bibitem [{\citenamefont {Barcel{\'o}}\ \emph {et~al.}(2011)\citenamefont
  {Barcel{\'o}}, \citenamefont {Liberati},\ and\ \citenamefont
  {Visser}}]{blv11}%
  \BibitemOpen
  \bibfield  {author} {\bibinfo {author} {\bibfnamefont {C.}~\bibnamefont
  {Barcel{\'o}}}, \bibinfo {author} {\bibfnamefont {S.}~\bibnamefont
  {Liberati}}, \ and\ \bibinfo {author} {\bibfnamefont {M.}~\bibnamefont
  {Visser}},\ }\bibfield  {title} {\enquote {\bibinfo {title} {Analogue
  gravity},}\ }\href@noop {} {\bibfield  {journal} {\bibinfo  {journal} {Living
  Rev. Relativity}\ }\textbf {\bibinfo {volume} {14}},\ \bibinfo {pages} {3}
  (\bibinfo {year} {2011})}\BibitemShut {NoStop}%
\bibitem [{\citenamefont {Bhattacharjee}(2017)}]{jkb17}%
  \BibitemOpen
  \bibfield  {author} {\bibinfo {author} {\bibfnamefont {J.~K.}\ \bibnamefont
  {Bhattacharjee}},\ }\bibfield  {title} {\enquote {\bibinfo {title} {Tunneling
  of the blocked wave in a circular hydraulic jump},}\ }\href@noop {}
  {\bibfield  {journal} {\bibinfo  {journal} {Phys. Lett. A}\ }\textbf
  {\bibinfo {volume} {381}},\ \bibinfo {pages} {733} (\bibinfo {year}
  {2017})}\BibitemShut {NoStop}%
\bibitem [{\citenamefont {{Lord Rayleigh}}(1914)}]{jws14}%
  \BibitemOpen
  \bibfield  {author} {\bibinfo {author} {\bibnamefont {{Lord Rayleigh}}},\
  }\bibfield  {title} {\enquote {\bibinfo {title} {On the theory of long waves
  and bores},}\ }\href@noop {} {\bibfield  {journal} {\bibinfo  {journal}
  {Proc. R. Soc. A}\ }\textbf {\bibinfo {volume} {90}},\ \bibinfo {pages} {324}
  (\bibinfo {year} {1914})}\BibitemShut {NoStop}%
\bibitem [{\citenamefont {Singha}\ \emph {et~al.}(2005)\citenamefont {Singha},
  \citenamefont {Bhattacharjee},\ and\ \citenamefont {Ray}}]{sbr05}%
  \BibitemOpen
  \bibfield  {author} {\bibinfo {author} {\bibfnamefont {S.~B.}\ \bibnamefont
  {Singha}}, \bibinfo {author} {\bibfnamefont {J.~K.}\ \bibnamefont
  {Bhattacharjee}}, \ and\ \bibinfo {author} {\bibfnamefont {A.~K.}\
  \bibnamefont {Ray}},\ }\bibfield  {title} {\enquote {\bibinfo {title}
  {Hydraulic jump in one-dimensional flow},}\ }\href@noop {} {\bibfield
  {journal} {\bibinfo  {journal} {Eur. Phys. J. B}\ }\textbf {\bibinfo {volume}
  {48}},\ \bibinfo {pages} {417} (\bibinfo {year} {2005})}\BibitemShut
  {NoStop}%
\bibitem [{\citenamefont {Mathur}\ \emph {et~al.}(2007)\citenamefont {Mathur},
  \citenamefont {DasGupta}, \citenamefont {Selvi}, \citenamefont {John},
  \citenamefont {Kulkarni},\ and\ \citenamefont {Govindarajan}}]{rama07}%
  \BibitemOpen
  \bibfield  {author} {\bibinfo {author} {\bibfnamefont {M.}~\bibnamefont
  {Mathur}}, \bibinfo {author} {\bibfnamefont {R.}~\bibnamefont {DasGupta}},
  \bibinfo {author} {\bibfnamefont {N.~R.}\ \bibnamefont {Selvi}}, \bibinfo
  {author} {\bibfnamefont {N.~S.}\ \bibnamefont {John}}, \bibinfo {author}
  {\bibfnamefont {G.~U.}\ \bibnamefont {Kulkarni}}, \ and\ \bibinfo {author}
  {\bibfnamefont {R.}~\bibnamefont {Govindarajan}},\ }\bibfield  {title}
  {\enquote {\bibinfo {title} {Gravity-free hydraulic jumps and metal
  femtoliter cups},}\ }\href@noop {} {\bibfield  {journal} {\bibinfo  {journal}
  {Phys. Rev. Lett.}\ }\textbf {\bibinfo {volume} {98}},\ \bibinfo {pages}
  {164502} (\bibinfo {year} {2007})}\BibitemShut {NoStop}%
\bibitem [{\citenamefont {Ray}\ \emph {et~al.}(2018)\citenamefont {Ray},
  \citenamefont {Sarkar}, \citenamefont {Basu},\ and\ \citenamefont
  {Bhattacharjee}}]{rsbb18}%
  \BibitemOpen
  \bibfield  {author} {\bibinfo {author} {\bibfnamefont {A.~K.}\ \bibnamefont
  {Ray}}, \bibinfo {author} {\bibfnamefont {N.}~\bibnamefont {Sarkar}},
  \bibinfo {author} {\bibfnamefont {A.}~\bibnamefont {Basu}}, \ and\ \bibinfo
  {author} {\bibfnamefont {J.~K.}\ \bibnamefont {Bhattacharjee}},\ }\bibfield
  {title} {\enquote {\bibinfo {title} {A theoretical prediction of rotating
  waves in {T}ype-{I} hydraulic jumps},}\ }\href@noop {} {\bibfield  {journal}
  {\bibinfo  {journal} {Phys. Lett. A}\ }\textbf {\bibinfo {volume} {382}},\
  \bibinfo {pages} {3399} (\bibinfo {year} {2018})}\BibitemShut {NoStop}%
\end{thebibliography}%
\end{document}